# Tipping the magnetic instability in paramagnetic $Sr_3Ru_2O_7$ by Fe impurities

M. Zhu[1], Y. Wang[2], P. G. Li[2], J. J. Ge[2], W. Tian[3], D. Keavney[4], Z. Q. Mao[2], X. Ke[1*]

[1]*Department of Physics and Astronomy, Michigan State University, East Lansing, Michigan 48824, USA*

[2]*Department of Physics and Engineering Physics, Tulane University, New Orleans, Louisiana 70118, USA*

[3]*Quantum Condensed Matter Division, Oak Ridge National Laboratory, Oak Ridge, Tennessee 37831, USA*

[4]*Advanced Photon Source, Argonne National Laboratory, Argonne, Illinois 60439, USA*

* corresponding author: ke@pa.msu.edu

We report the magnetic and electronic properties of the bilayer ruthenate $Sr_3Ru_2O_7$ upon Fe substitution for Ru. We find that $Sr_3(Ru_{1-x}Fe_x)_2O_7$ shows a spin-glass-like phase below 4 K for $x = 0.01$ and commensurate E-type antiferromagnetically ordered insulating ground state characterized by the propagation vector $\mathbf{q}_c = (0.25\ 0.25\ 0)$ for $x \geq 0.03$, respectively, in contrast to the paramagnetic metallic state in the parent compound with strong spin fluctuations occurring at wave vectors $\mathbf{q} = (0.09\ 0\ 0)$ and $(0.25\ 0\ 0)$. The observed antiferromagnetic ordering is quasi-two-dimensional with very short correlation length along the $c$ axis, a feature similar to the Mn-doped $Sr_3Ru_2O_7$. Our results suggest that this ordered ground state is associated with the intrinsic magnetic instability in the pristine compound, which can be readily tipped by the local magnetic coupling between the $3d$ orbitals of the magnetic dopants and Ru $4d$ orbitals.

Understanding the complex behaviors of strongly correlated electron systems is a central challenge in condensed matter physics. Prototypical phenomena, such as unconventional superconductivity [1, 2], colossal magnetoresistance [3, 4], and multiferroicity [5, 6], have been explored intensely both theoretically and experimentally for decades. Ruddlesden-Popper series layered perovskite ruthenates $(Sr,Ca)_{n+1}Ru_nO_{3n+1}$ are intriguing material systems where a diversity of fascinating phenomena have been discovered, including unconventional $p$-wave spin-triplet superconductivity in the single-layer $Sr_2RuO_4$ ($n$ = 1) [7-9], Mott insulating state in $Ca_2RuO_4$ [10], and bulk spin valve effect in $Ca_3Ru_2O_7$ [11, 12]. The bilayer $Sr_3Ru_2O_7$ is another very interesting compound, which exhibits a paramagnetic metallic ground state and is close to ferromagnetic instability [13]. More intriguingly, while the system shows Fermi liquid behavior at zero field, it possesses a magnetic-field-tuned quantum critical point (QCP), where non-Fermi liquid behavior [14] and highly anisotropic magnetoresistance [15] emerge. Recently neutron diffraction measurements have identified two different magnetically ordered phases close to the QCP [16]. Furthermore, it is revealed that ferromagnetic and antiferromagnetic spin fluctuations coexist in this system [17], and the latter is suggested to be dominant near the critical field, which is unexpected for a metamagnetic transition [18]. In general, the strong interplay among charge, spin, orbital, and lattice degrees of freedom in these systems often gives rise to emergent phenomena that can be readily tuned by external stimuli [19, 20].

Chemical doping is an effective approach to tailor materials' properties by stabilizing one of the competing phases while suppressing others. For instance, in $(Sr,Ca)_{n+1}Ru_nO_{3n+1}$ the isovalent Ca substitution for Sr leads to structural distortions, which tends to drive the system towards antiferromagnetism and non-metallicity [21, 22]. On the other hand, doping 3$d$ transition-metal elements into Ru sites can induce very distinct effects on the magnetic and electronic properties that strongly depend on the choice of 3$d$ dopants and the parent ruthenate compounds [23]. In Sr$_2$RuO$_4$, a very small amount of nonmagnetic Ti [24] or magnetic Mn impurities [25] suppress the antiferromagnetic spin fluctuations that originate from Fermi surface nesting at $\mathbf{q}_{ic}$ = (0.3 0.3 $L$) in the parent compound [26], and give rise to a short-range, static incommensurate spin density wave order (SDW) characterized by the same wave vector [25, 27]. In contrast, Fe doping leads to a commensurate SDW state with a different propagation vector $\mathbf{q}_c$ = (0.25 0.25 0), whereas the dominant magnetic correlations are still centered at the incommensurate wave vector $\mathbf{q}_{ic}$ [28]. Similarly, in Ca$_3$Ru$_2$O$_7$, Ti and Mn dopants can drive the system into a G-type antiferromagnetic Mott insulating ground state [29-31], while Fe doping results in a localized state accompanied with an incommensurate magnetic order [32].

However, very intriguingly, Ti and Mn dopants have different effects on the physical properties of Sr$_3$Ru$_2$O$_7$. While both Ti and Mn doping lead to insulating electronic transport behavior at low temperature, doping Ti gives rise to an incommensurate SDW order with the propagation wave vector $\mathbf{q}_{ic}$ = (0.24 0.24 0) [33, 34] but Mn-doped Sr$_3$Ru$_2$O$_7$ exhibits [35, 36] a commensurate E-type antiferromagnetic order characterized by $\mathbf{q}_c$ = (0.25 0.25 0) [37]. In both cases, the magnetic orders do not reflect the dominant magnetic correlations at $\mathbf{q}$ = (0.09 0 0) and (0.25 0 0) in the pristine Sr$_3$Ru$_2$O$_7$ [17], in contrast to those in Ti- and Mn-doped Sr$_2$RuO$_4$ [25, 27]. In addition, synchrotron x-ray absorption spectroscopy measurements have found that Mn dopants in Sr$_3$Ru$_2$O$_7$ possess a valence value of Mn$^{3+}$, and show an inversion of the conventional crystal-field level hierarchy, which is suggested to be due to the hybridization between Ru-O 4$d$-2$p$ bands and Mn 3$d$ orbitals [38]. Thus, one can see that the physical properties of ruthenates are very susceptible to 3$d$ transition-metal dopants and novel ground states often emerge upon chemical substitution. Considering the remarkably distinct doping effects induced by Fe and Mn/Ti in both Sr$_2$RuO$_4$ and $Ca_3Ru_2O_7$ systems, it would be interesting to study how Fe doping affects the ground state properties of $Sr_3Ru_2O_7$.

In this paper, we report the magnetic and electronic properties of Fe-doped $Sr_3Ru_2O_7$. In contrast to the paramagnetic metallic state in the parent compound, $Sr_3(Ru_{1-x}Fe_x)_2O_7$ ($x = 0.01$) shows a metallic spin-glass-like ground state, whereas for $x = 0.03$ an insulating phase with quasi-two-dimensional E-type antiferromagnetic order characterized with the propagation vector $\mathbf{q}_c$ = (0.25 0.25 0) is observed below $T_N \sim 40$ K. These features are similar to the doping effects in Mn-doped $Sr_3Ru_2O_7$ [37], which suggest that the induced ordered state upon Fe and Mn doping originates from the intrinsic instability of $Sr_3Ru_2O_7$.

The single crystals of $Sr_3(Ru_{1-x}Fe_x)_2O_7$ ($x = 0.01, 0.03$) were grown using the floating zone technique. Magnetization, specific heat, and resistivity measurements were performed using the Physical Property Measurement System (PPMS, Quantum Design). Neutron diffraction experiments were carried out using the HB-1A thermal neutron triple-axis spectrometer at High Flux Isotope Reactor in Oak Ridge National Laboratory. The energy of the incident neutrons is fixed as $E_i = 14.6$ meV. The single-crystal samples were oriented in ($H\,K\,0$) and ($H\,H\,L$) scattering planes, where $H$, $K$, $L$ are in reduced lattice units (r.l.u.) $2\pi/a$, $2\pi/b$, and $2\pi/c$, and were mounted in an aluminum sample can and cooled down using a closed-cycle helium refrigerator down to 4 K. For convenience, we describe our neutron diffraction data using the tetragonal lattice symmetry $I4/mmm$, with $a = b = 3.874$ Å and $c = 20.69$ Å. The neutron intensity was presented in the unit of counts per monitor count unit (mcu), with 1 mcu corresponding to approximately ~1 s. Synchrotron x-ray absorption spectroscopy experiments were performed using the beamline 4-ID-C at Advanced Photon Source in Argonne National Laboratory to measure the valence state of Fe dopants.

The main panels of Figure 1(a) and 1(b) show the temperature dependence of the magnetic susceptibility $\chi_c$ of $Sr_3(Ru_{1-x}Fe_x)_2O_7$ ($x$ = 0.01, 0.03) after zero-field-cooling (ZFC) and field-cooling (FC), respectively. In the pristine $Sr_3Ru_2O_7$, the magnetic susceptibility is nearly isotropic with a pronounced peak at $T$ = 16 K in both $\chi_c$ and $\chi_{ab}$ [13], which is ascribable to the crossover of the nature of the dominant magnetic fluctuations from ferromagnetic to antiferromagnetic upon cooling [17]. No hysteresis effect is observed between the ZFC and FC data [13]. In contrast, for $x$ = 0.01, the peak at 16 K is completely suppressed in both $\chi_c$ and $\chi_{ab}$. Instead, $\chi_c$ (T) shows a maximum at $T_g \sim 4$ K below which a bifurcation between the ZFC and FC data is clearly seen [Fig. 1(a)], characteristic of a spin-glass-like state. Figure 1(c) displays the magnetization as a function of the magnetic field applied along the $c$ axis at $T$ = 2 and 6 K, respectively. The hysteretic behavior observed at 2 K indicates that short-range ferromagnetic correlations develop between the neighboring $RuO_2$ layers along the $c$ axis below $T_g$. On the contrary, as shown in the inset of Fig. 1(a), the in-plane magnetic susceptibility $\chi_{ab}$ shows paramagnetic behavior down to 2 K and there is no bifurcation between the ZFC and FC curves. These results suggest that Fe doping in $Sr_3Ru_2O_7$ induces strong magnetic anisotropy with the easy axis along the $c$ direction, similar to Fe-doped $Sr_2RuO_4$ [28]. Furthermore, the metamagnetic transition observed in the parent compound has been completely smeared out, as shown in the inset of Fig. 1(c), which implies that Fe doping drives the system away from metamagnetism, as reported in Ti-doped $Sr_3Ru_2O_7$ [33].

Interestingly, for $x$ = 0.03 a sharp peak in magnetic susceptibility develops in both $\chi_c$ and $\chi_{ab}$ at $T_N \sim 40$ K, suggestive of an onset of paramagnetic-antiferromagnetic phase transition. Neither the bifurcation between the magnetic susceptibility curves measured with ZFC and FC histories [Fig. 1(b)] nor the hysteresis in the isothermal magnetization data [Fig. 1(d)] is observed. In addition, it is noteworthy that at 9 T the magnetization at 2 K is much smaller than that at 50 K.

These results suggest that the nature of the peaks in the magnetic susceptibility data are fundamentally different from that in the $x = 0.01$ compound, which implies that 3% Fe dopants lead to the formation of a long-range antiferromagnetic ordering.

Figure 2(a) shows the temperature dependence of the specific heat for Fe-doped $Sr_3Ru_2O_7$ ($x = 0.01$ and $0.03$). It is worth pointing out two distinctive features. First, for the $x = 0.03$ compound, an anomaly is clearly observed close to $T_N \sim 40$ K, which is indicative of the formation of a long-range antiferromagnetic order. In contrast, no anomaly in specific heat is observed in the $x = 0.01$ compound in this temperature range. Second, the $x = 0.01$ compound exhibits an upturn below $T \sim 10$ K, which is presumably of magnetic origin. Note that $C_p/T$ measured at the lowest temperature ($\sim 0.21$ J / K$^2$ mol) is much larger than that for the $x = 0.03$ compound ($\sim 0.06$ J / K$^2$ mol). Similar behaviors have been observed in Ti-doped $Sr_3Ru_2O_7$ [34], which has been suggested to be associated with the spin fluctuations. The increase of the specific heat in the low temperature limit in the disordered state (e.g., $x = 0.01$) can be ascribed to the softening of the magnetic fluctuations when being closer to the magnetically ordered state induced upon doping; whereas the suppression of the value of $C_p$/T in the antiferromagnetically ordered phase for the $x = 0.03$ compound is due to the opening of the gap in the spin excitation spectrum [34].

The temperature dependence of the in-plane resistivity $\rho_{ab}$ of Fe-doped $Sr_3Ru_2O_7$ is presented in Fig. 2(b). Similar to the pristine $Sr_3Ru_2O_7$, the $x = 0.01$ compound exhibits metallic behavior down to 2 K. At low temperature, $\rho_{ab}(T)$ shows $T^2$ dependence as shown in the inset, which is characteristic of Fermi liquid behavior. However, the field-induced anisotropic, highly resistive state in the parent compound [15] is completely suppressed in the $x = 0.01$ compound, consistent with the absence of the metamagnetic transition in the magnetic susceptibility data shown in the inset of Fig. 1(c). In contrast, a metal-insulator transition (MIT) is observed in the $x = 0.03$ compound at $T_{MIT} \sim 44$ K. Similar behaviors have been observed in Ti- and Mn-doped $Sr_3Ru_2O_7$ [34, 36]. It has been suggested that the MIT in the Mn-doped compound is Mott type, induced by electron correlations instead of Slater type due to the formation of the antiferromagnetic order [35]. As $T_{MIT}$ of the $x = 0.03$ compound of Fe-doped $Sr_3Ru_2O_7$ is very close to the antiferromagnetic transition temperature $T_N \sim 40$ K, it might be helpful to study this material system with higher Fe doping concentrations to resolve the nature of the MIT in this compound.

In order to determine the spin structure of the antiferromagnetically ordered phase in Fe-doped $Sr_3Ru_2O_7$ ($x = 0.03$), we have carried out neutron diffraction measurements. The magnetic Bragg peaks were observed at $\mathbf{q}_c = (0.25\ 0.25\ 0)$ and the equivalent positions in the reciprocal space, such as (0.75 0.75 0) and (1.25 1.25 0), etc. Figure 3(a) shows the neutron diffraction scans along the [1 1 0] direction across $\mathbf{q}_c$ at representative temperatures. The data are well fitted by Gaussian functions and the peak width is resolution limited, indicating a long-range ordering in the basal plane (*ab* plane). Temperature dependence of the peak intensity of $\mathbf{q}_c$ is shown in Fig. 3(b). The peak intensity starts to increase at $T \sim 40$ K on cooling, consistent with $T_N$ obtained by the magnetic susceptibility measurements. Fig. 3(c) shows the scans along the [0 0 1] direction across $\mathbf{q}_c$ at selected temperatures. These curves are best fitted by Lorentzian functions, in contrast to Gaussian for scans along the [1 1 0] direction, which indicates short-range magnetic correlations (1/FWHM $\sim 1.2c$, $c = 20.69$ Å, at $T = 4$ K) along the $c$ axis. These results suggest that the magnetic ordering is quasi-two-dimensional, similar to that in Ti- and Mn-doped $Sr_3Ru_2O_7$ [34, 37]. Figure 3(d) displays the same scan across $\mathbf{q}_c = (0.25\ 0.25\ 0)$ and (0.75 0.75 0) for a wider $L$ range at $T = 4$ K. It is worth noting that the magnetic intensities are only observed at even values of $L$. Furthermore, while the intensity of (0.25 0.25 $L$) with $L = 0$ and 2 is much greater than that of (0.75 0.75 $L$), at $L = 4$ and 6 the intensity of these two types of magnetic Bragg peaks are comparable.

Possible magnetic structure models have been explored by the representation analysis using FullProf [39] and the magnetic symmetry analysis using the programs at Bilbao Crystallographic Server [40]. The obtained results are in agreement with each other. Due to the orthorhombic crystal symmetry and the propagation vector $\mathbf{q}_c$ = (0.25 0.25 0) (i.e., (0.5 0 0) in orthorhombic symmetry notation [37]), our data are best described by the E-type antiferromagnetic structure with zigzag spin chains in the basal plane, as shown in Fig. 4(a). The magnetic moments in a bilayer are ferromagnetically aligned, similar to that in Mn-doped $Sr_3Ru_2O_7$ [37]. For this magnetic structure model, the squared structure factors of the magnetic reflections $\mathbf{q}_c$ = (0.25 0.25 $L$) and (0.75 0.75 $L$) are nonzero only for even values of $L$:

$$|S(0.25\ 0.25\ L)|^2 = |S(0.75\ 0.75\ L)|^2 \sim \cos^2(2\pi\Delta L),$$

where $2\Delta \approx 0.20$ is the separation between neighboring $RuO_2$ planes. This is in line with the data shown in Fig. 3(d). In addition, the comparison of the intensity of (0.25 0.25 $L$) and (0.75 0.75 $L$) [Fig. 3(d)] also provides clues on the spin orientation. The cross section of magnetic neutron diffraction is given by

$$\sigma(\mathbf{q}) \propto |F(\mathbf{q})|^2 [1 - (\hat{\mathbf{q}} \cdot \widehat{\mathbf{M}})^2] |S(\mathbf{q})|^2,$$

where $F(\mathbf{q})$ is the magnetic form factor, $S(\mathbf{q})$ is the structure factor, and $\hat{\mathbf{q}}$, $\widehat{\mathbf{M}}$ the unit vectors of the wave vector $\mathbf{q}$ and magnetic moment $\mathbf{M}$. The fact that the intensity of (0.25 0.25 $L$) and (0.75 0.75 $L$) for $L$ = 4 is comparable suggests that the magnetic moments are along the $c$ axis such that the polarization factor $[1 - (\hat{\mathbf{q}} \cdot \widehat{\mathbf{M}})^2]$ is greater for the latter, since the structure factors are equal for these two reflections and the magnetic form factors decrease rapidly with the increasing modulus of **q**. Similar argument holds for the $L$ = 6 case. The ordered moment is estimated to be

~ 0.5 $\mu_B$ for the $x = 0.03$ compound, which is much smaller than ~2 $\mu_B$ expected for fully localized $Ru^{4+}$ magnetic moments.

There are several remarkable features observed in Fe-doped $Sr_3Ru_2O_7$ that may help to elucidate the effects of 3$d$ transition-metal dopants on this system. First of all, the magnetic structure in Fe-doped $Sr_3Ru_2O_7$ is very similar to that in the Mn-doped compound, which is independent on the Mn doping concentration [36]. This suggests that the magnetic ordering is an inherent instability of the system and the role of Fe and Mn dopants is to tip the balance between the competing magnetic tendencies (see discussion below). Second, the propagation wave vector of the magnetic order induced by 3$d$ transition-metal doping, e.g., Ti [34], Mn [36], and Fe in this study, does not correspond to the dominant spin fluctuations in the pristine $Sr_3Ru_2O_7$, which are centered on the principal axes of the tetragonal Brillouin zone at $\mathbf{q} = (0.09\ 0\ 0)$ and $(0.25\ 0\ 0)$ [17]. It has been reported that these incommensurate antiferromagnetic fluctuations originate from the nesting of the Fermi surface [17], and that in the single-layer $Sr_2RuO_4$, Ti and Mn doping can suppress the spin fluctuations and stabilize a spin density wave ordering with a propagation vector the same as the nesting wave vector of the Fermi surface [25, 27]. Therefore, the observation of a commensurate E-type magnetic ordering with the propagation vector distinct from the wave vectors of the spin fluctuations in the pristine $Sr_3Ru_2O_7$ suggests that the impurity-induced magnetic ordering in this bilayer system is not due to Fermi surface nesting, a feature different from that in the Ti- and Mn-doped $Sr_2RuO_4$ [25, 27].

Third, it is well known that the magnetism in layered ruthenates is strongly correlated with structural distortions. In Mn-doped $Sr_3Ru_2O_7$, a change in the lattice constants of ~0.1% has been reported at $T_N$ by x-ray diffraction measurements [36]. However, no change in the lattice constants at either $T_N$ or $T_{MIT}$ was convincingly observed in Fe-doped $Sr_3Ru_2O_7$ via single-crystal neutron diffraction measurements, although we cannot exclude the possibility that the change is too small to be seen within the instrumental resolution. In the magnetically ordered state, it has been reported that Mn doping reduces the rotation of the RuO$_6$ octahedron and enhances the octahedral flattening via a reduction in the apical Ru-O bond length [41]. We have performed synchrotron x-ray absorption measurements on the $x = 0.03$ compound at the Fe $L_{2,3}$ edge at room temperature, together with two reference samples with well-defined valence values. As shown in Fig. 4(b), the peak position and the line shape of the data suggest that the Fe dopants are in the Fe$^{3+}$ valence state, the same as Mn$^{3+}$ in Mn-doped Sr$_3$Ru$_2$O$_7$ [38]. Since the ionic radii of Fe$^{3+}$ and Mn$^{3+}$ are the same and larger than that of Ru$^{4+}$ [42], we expect that Fe dopants give rise to similar structural effects as Mn dopants. Theoretical studies on the single-layer compound Ca$_{2-x}$Sr$_x$RuO$_4$ have revealed that the rotation of the RuO$_6$ octahedron favors ferromagnetism and the subsequent tilting leads to antiferromagnetism, while the flattening of the octahedron is essential to stabilize the FM or AFM state [21]. However, the observed E-type antiferromagnetic order in Fe- and Mn-doped Sr$_3$Ru$_2$O$_7$ is not in line with this picture that the magnetic order is due to the doping-induced changes in the structural distortion discussed above.

Finally, the interplay between the 3$d$ orbitals of the dopants and the Ru 4$d$ orbitals might be essential to the impurity-induced magnetic ordering and MIT. For instance, it has been revealed that the hybridization between Mn 3$d$ and Ru-O orbitals can reverse the conventional hierarchy of the crystal-field levels of the half-filled Mn $e_g$ orbital in Mn-doped Sr$_3$Ru$_2$O$_7$ [38]. A recent study on the single-layer Fe-doped Sr$_2$RuO$_4$ has found a commensurate magnetic ordering characterized by the propagation wave vector (0.25 0.25 0), and density functional theory calculations showed that Fe doping barely changes the Fermi surface but induces strong back-polarization on the Ru neighboring Fe which may lead to the commensurate magnetic order [28]. Considering the similarity between the electronic structures of $Sr_2RuO_4$ and $Sr_3Ru_2O_7$ [38], it is very likely that a similar scenario holds true for the Fe-doped $Sr_3Ru_2O_7$, though first-principles calculations are highly warranted to examine this conjecture.

In summary, we have investigated the magnetic and electronic properties of the bilayer ruthenate $Sr_3Ru_2O_7$ doped by Fe. We find that in contrast to the paramagnetic Fermi liquid ground state in the pristine compound, 1% Fe substitution leads to a metallic spin-glass-like state, whereas an insulating, E-type antiferromagnetically ordered phase is induced below $T_N \sim 40$ K for the 3% Fe-doped compound. This magnetic ordering is quasi-two-dimensional with short correlation length along the $c$ direction, similar to the ground state observed in Mn-doped $Sr_3Ru_2O_7$, suggesting that the induced ordered state is associated with the intrinsic magnetic instability of the pristine compound which can be readily tuned via the local coupling between magnetic dopants and Ru hosts.

Work at Michigan State University was supported by the National Science Foundation under Award No. DMR-1608752 and the start-up funds from Michigan State University. Work at Tulane University was supported by the U.S. Department of Energy (DOE) under EPSCOR Grant No. DE-SC0012432 with additional support from the Louisiana Board of Regents (support for crystal growth). Work at ORNL's HFIR was supported by the Scientific User Facilities Division, Office of Basic Energy Sciences, DOE. This research used resources of the Advanced Photon Source, a U.S. Department of Energy Office of Science User Facility operated for the DOE Office of Science by Argonne National Laboratory under Contract No. DE-AC02-06CH11357.

**Figure Captions**

Figure 1. (a),(b) Temperature dependence of the magnetic susceptibility $\chi_c$ of $Sr_3(Ru_{1-x}Fe_x)_2O_7$ ($x$ = 0.01 and 0.03) after ZFC and FC. Insets show the in-plane magnetic susceptibility $\chi_{ab}$ as a function of temperature after ZFC and FC. (c),(d) Isothermal magnetization as a function of the magnetic field of $Sr_3(Ru_{1-x}Fe_x)_2O_7$ ($x$ = 0.01 and 0.03), respectively. The field is applied along the $c$ axis. Inset shows the data of the $x$ = 0.01 compound at 2 K up to 9 T.

Figure 2. (a) Specific heat of $Sr_3(Ru_{1-x}Fe_x)_2O_7$ as a function of temperature for $x$ = 0.01 and 0.03. The blue arrow denotes the Neel temperature $T_N$ obtained from the magnetic susceptibility measurements. (b) In-plane resistivity $\rho_{ab}$ of $Sr_3(Ru_{1-x}Fe_x)_2O_7$ as a function of temperature. Inset shows $\rho_{ab}$ vs $T^2$ in the low temperature regime for $x$ = 0.01. The red line is a fit using linear function.

Figure 3. Neutron diffraction data of $Sr_3(Ru_{1-x}Fe_x)_2O_7$ ($x$ = 0.03). (a) Radial scans around $\mathbf{q}_c$ = (0.25 0.25 0) along the [1 1 0] direction at representative temperatures. $H$ is in reduced lattice unit. (b) Temperature dependence of the peak intensity of $\mathbf{q}_c$ = (0.25 0.25 0) showing the magnetic ordering at $T_N \sim$ 40 K. (c),(d) Scans around $\mathbf{q}_c$ = (0.25 0.25 0) and (0.75 0.75 0) along the [0 0 1] direction, respectively. $L$ is in reduced lattice unit.

Figure 4. (a) Schematic of the E-type antiferromagnetic order of one bilayer in $Sr_3(Ru_{1-x}Fe_x)_2O_7$ ($x$ = 0.03). (b) Synchrotron x-ray absorption measurements on Fe-doped $Sr_3Ru_2O_7$ ($x$ = 0.03) and the reference samples FeO ($Fe^{2+}$) and $Fe_2O_3$ ($Fe^{3+}$).

**Figure 1**

M. Zhu et al.

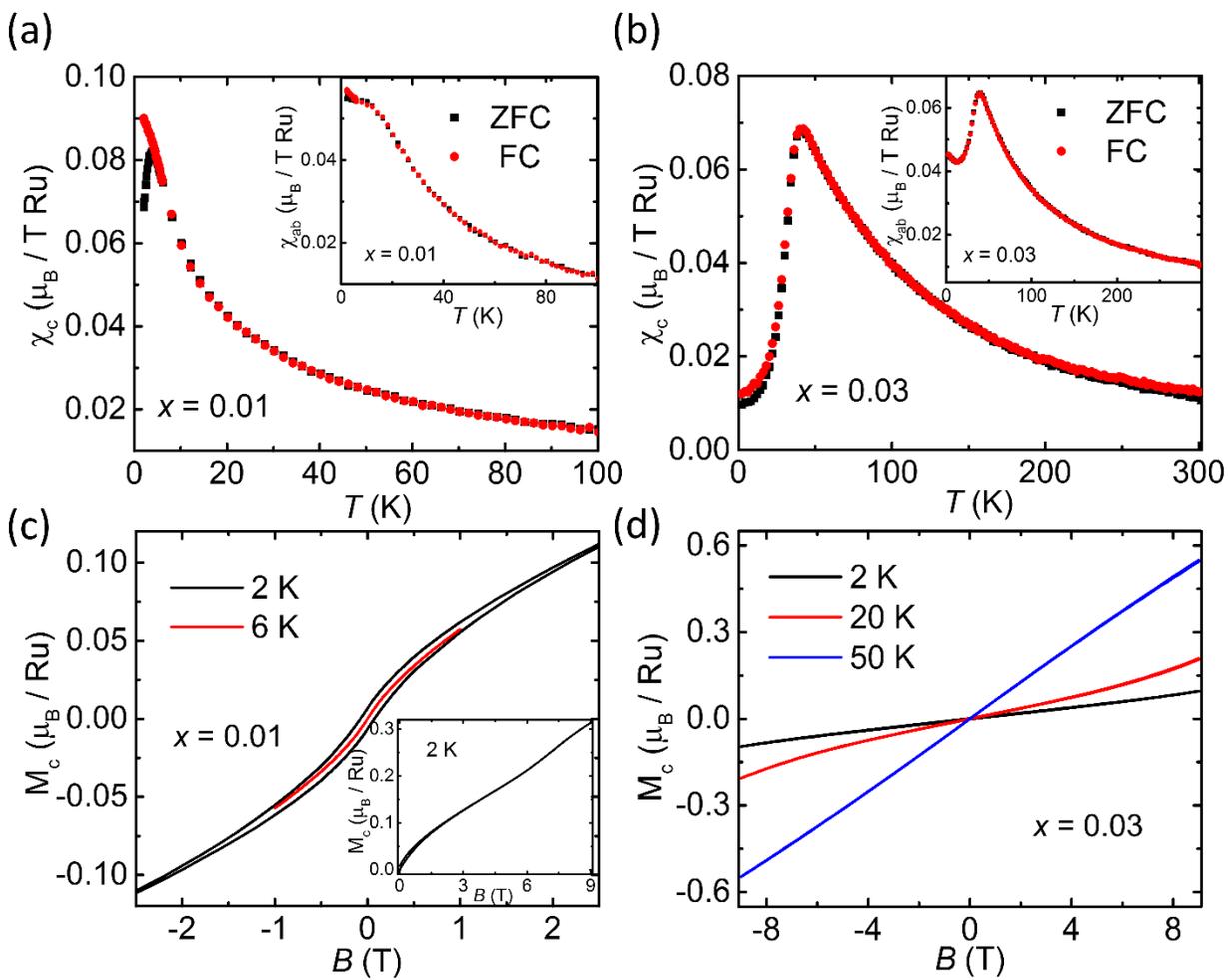

**Figure 2**

M. Zhu et al.

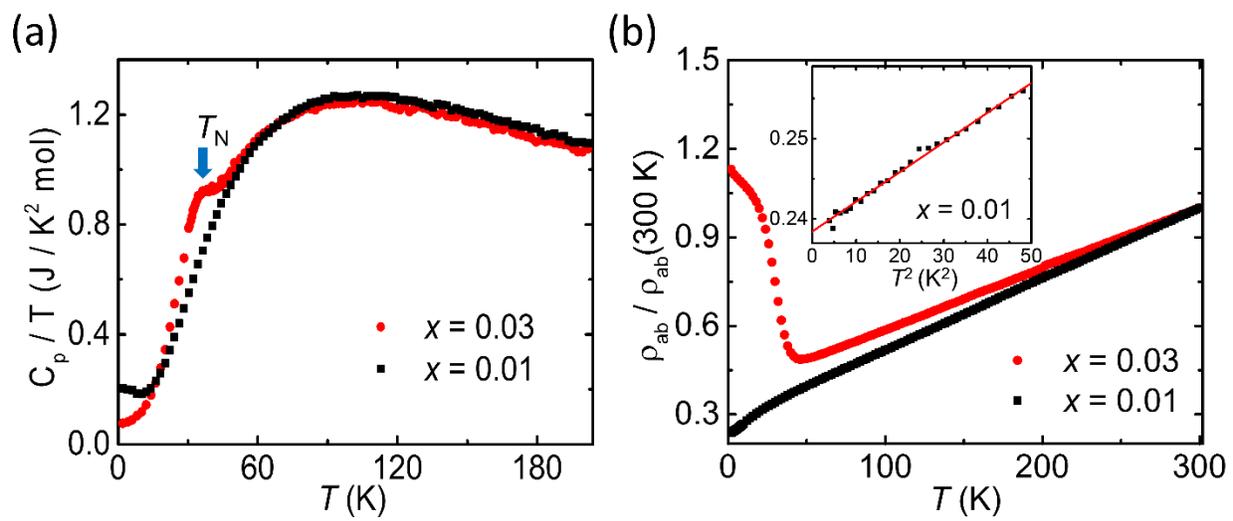

**Figure 3**

M. Zhu et al.

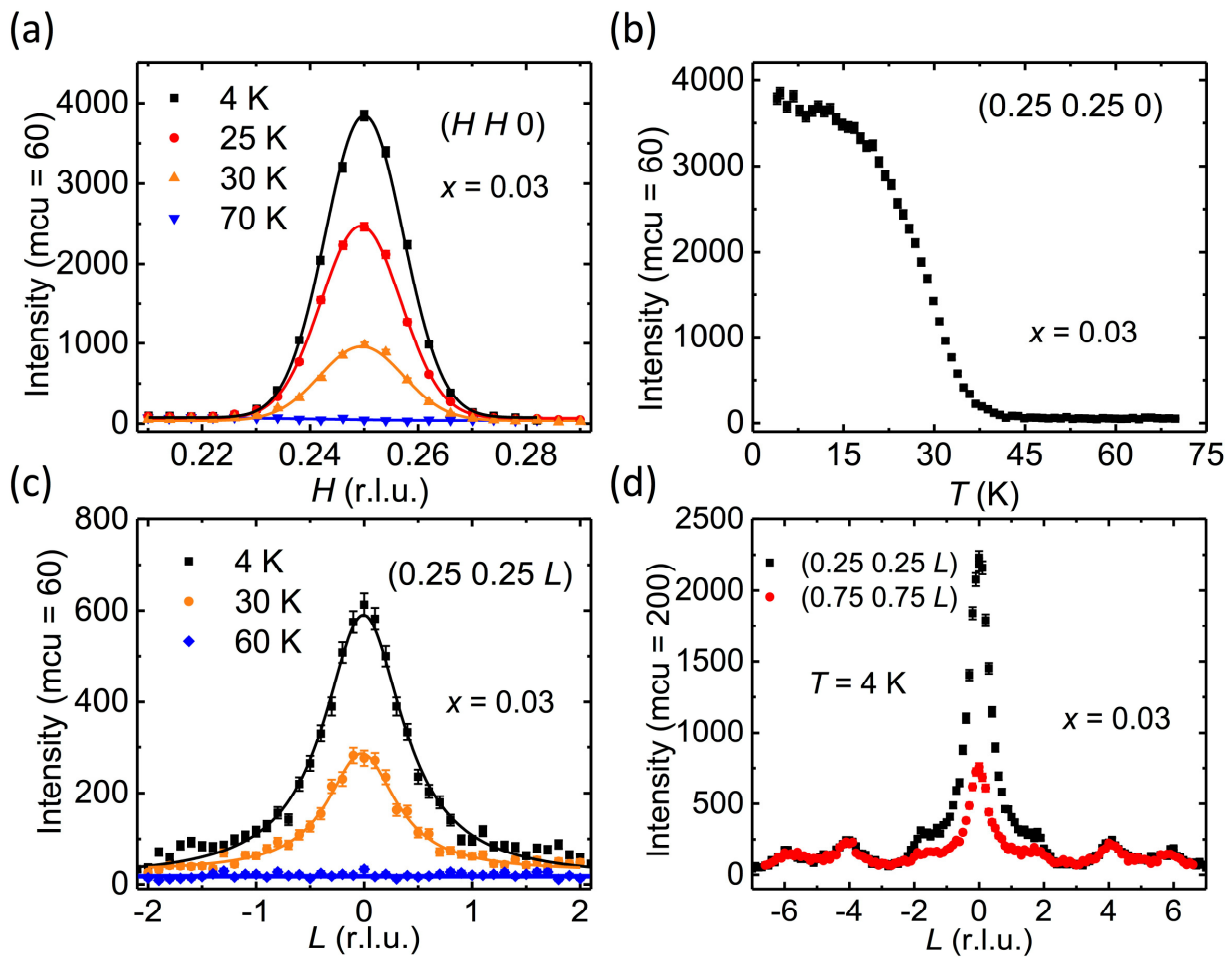

**Figure 4**

M. Zhu et al.

(a) 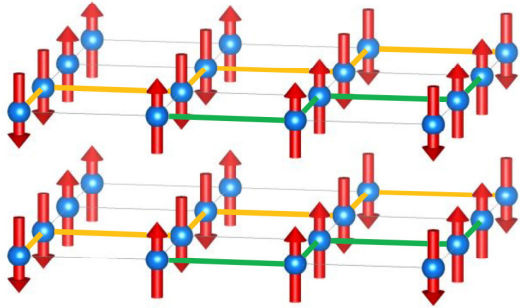
(b) 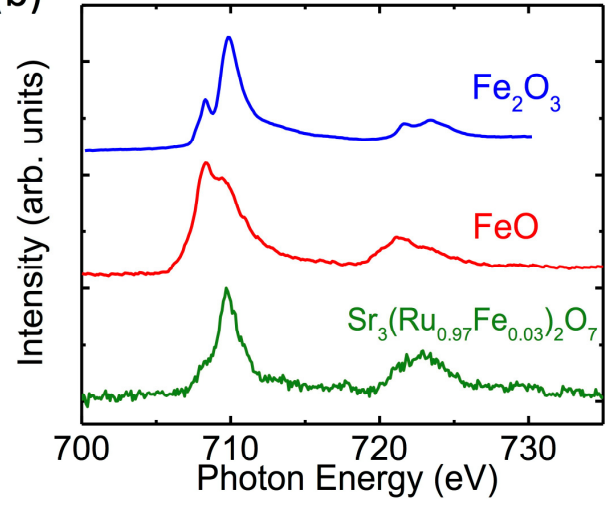